\documentclass{elsart}
\journal{New Astronomy}
\usepackage{epsfig}

\def\url#1{{\ttfamily\def\/{/\discretionary{}{}{}}#1}}

\begin{document}

\begin{frontmatter}
\title{CMB Observations: improvements of the performance of correlation radiometers
by signal modulation and synchronous detection}
\author[mibic]{Daniele Spiga\thanksref{fn2}},
\author[mibic]{Elia Battistelli\thanksref{now}},
\author[mibic]{Giuliano Boella\thanksref{fn1}},
\author[mibic]{Massimo Gervasi\thanksref{fn1}},
\author[mibic]{Mario Zannoni\thanksref{fn1}},
\author[mibic]{Giorgio Sironi\thanksref{fn1}}
\thanks[fn1]{E-mail: name.surname@mib.infn.it}
\thanks[now]{Present address: Physics Dept. - University of Rome La Sapienza, 
~E-mail: Elia.Stefano.Battistelli@roma1.infn.it}
\thanks[fn2]{E-mail: daniele.spiga@tin.it}
\address[mibic]{Dipartimento di Fisica G. Occhialini - University 
of Milano Bicocca - Milano -Italy}
% use the thanksref command within \title, \author or \address for footnotes:
% \title{\thanksref{label1}}
% \thanks[label1]{}
% \author{\thanksref{label2}}
% \thanks[label2]{}
% \address{\thanksref{label3}}
% \thanks[label3]{}
% including your email address
% \address{\thanksref{email}}
% \thanks[email]{E-mail: }

\begin{abstract}
Observation of the fine structures (anisotropies, polarization, 
spectral distortions) of the Cosmic Microwave Background (CMB) is hampered 
by instabilities, $1/f$ noise and asymmetries of the radiometers used to carry
on the measurements. 
Addition of modulation and synchronous detection allows to increase the 
overall stability and the noise rejection of the radiometers used
for CMB studies. In this paper we discuss the advantages this technique has
when we try to detect CMB polarization. The behaviour of a two channel 
correlation receiver to which phase modulation and synchronous detection
have been added is examined.  Practical formulae for evaluating the 
improvements are presented.
\end{abstract}

\begin{keyword}
% keywords here, in the form keyword \sep keyword
% PACS code here, in the form \PACS code \sep code
Cosmic Background Radiation, Radiometers, Polarimeter, Correlation
\PACS:03.09.05, 03.19.1, 12.03.1
\end{keyword}
\end{frontmatter}

% main text
\begin{section}{Introduction}
The fine structures (spatial anisotropies, spectral distortions, 
residual polarization) of the Cosmic Microwave Background
(CMB), relic of the Big Bang, are among the most powerfull tools available for 
probing the evolution of the Universe (for a general discussion 
see for instance \cite{Part95} and \cite{stag1}). Their detection
can in fact be used to go back at least up to 
redshisft $Z \simeq 10^6 ~-~ 10^7$, when the Universe was extremely young and the 
matter condensations and objects we observe today not yet formed. So far however
only spatial anisotropies have been discovered (\cite{Smoo92}) and are currently
studied (e.g. \cite{debe00}, \cite{hana00}). Spectral distortions and polarization 
escaped so far detection and only upper limits to their amplitude have 
been obtained (see for instance \cite{Siro99}, \cite{Siro01}, \cite{stag0}). 
In fact the expected signals are extremely faint when compared with spurious effects
produced by small instabilities of the receiver, $1/f$ noise, 
pick up of tiny fractions of undesired signals, deviations of the system components from
their ideal behaviour etc. Therefore many radiometers succesfully used for standard 
radioastronomical observations become useless when applied to search for the CMB 
fine structures: {\it ad hoc} systems are necessary. For polarization studies
correlation receivers are usually preferred 
(e.g. \cite{Siro98}, ~\cite{MAP98}, ~\cite{Sport}, ~\cite{Torb99}, ~\cite{Hedm01}, ~\cite{Keat01})
because they make possible simultaneous measurements of pair of Stokes parameteres, 
(U and Q or U and V), and in principle allow to detect signals of few 
$\mu K$. Unfortunately such a sensitivity is barely sufficient because the expected
amplitude of the CMB polarized component is probably below the $\mu K$ level. Therefore
the receivers so far used for studies of the CMB polarization have to be improved.
   In the following we discuss the limits of a standard two channel
correlation receiver and the improvements in noise rejection and offset
cancellation one obtains adding phase modulation (at the system
front end) and synchronous detection (at the back end).

\end{section}

\begin{section}{Radiometers and noise}

A radiometer (see figure 1)~is a chain of 
linear devices plus a square law detector which amplify and convert
the signal $s(t)$, collected by the antenna, and the noise $n(t)$,
produced by the system components, into DC signals $V_s(t)$ and $V_n(t)$ 
proportional to the power content of $s$ and $n$

\begin{equation}
    V(t) = [|s(t)|^2~+~|n(t)|^2]~G = [v_s(t)+ v_n(t)]~G = V_s(t)~+~V_n(t)
\label{eq:radiometro}
\end{equation}
Here the power gain G includes the detector responsivity, (conversion
factor between power and output voltage (current)), 
$s$ and $n$ are electric (magnetic) fields with zero 
mean values, while $V_i$ and $v_i$ are voltages, proportional to the power 
content of the signals, whose mean value is greater than zero: 
all of them fluctuate and behave as noise \cite{Vand54}.
 
Let's call 
$x(t)$, ~$E\{x(t)\}$ and $\mu_x = E\{x(t)\}$ one of these signals, its 
expectation value and its mean value respectively. The signal variance 
is (e.g. \cite{Vand54}, \cite{Rolf86}):

\begin{equation}
	\sigma_x^2 = E\{x^2\}~-~E^2\{x\} = \int^\infty _0
w_x(\nu) \, d\nu  - \mu_x^2
\label{eq:sigma}
       \end{equation}

where $w_x(\nu)=|a(\nu)|^2$ is the signal power spectrum and
\begin{equation}
 a(\nu) = \int^{+\infty} _{-\infty} x(t) e^{-2\pi\nu t} dt
\label{eq:fourier}
\end{equation}
the Fourier transform of $x(t)$.
In practice we can write

\begin{equation}
\sigma_x^2~+~\mu_x^2 =  \int^\infty _0
w_x(\nu) \, d\nu  ~\simeq~ \int^{\nu_{max}}_{\nu_{min}}
w_x(\nu) \, d\nu  =  \int^{1/\tau}_{1/T}
w_x(\nu) \, d\nu 
\label{eq:sigmammu} 	
       \end{equation}

where $\nu_{min} \simeq 1/T$ and $\nu_{max} \simeq 1/\tau$ are the minimum and maximum 
frequencies of the signal fluctuations accepted by the system, $\tau$ the sample collecting
time, $N$ the number of samples and $T = N\tau$ the total observing time. To guarantee 
that the samples are statistically independent $\tau$ must be longer
than the system time constant $\tau'$ ($\tau \geq 3~\tau'$). 

\begin{figure}
          \begin{center}
\includegraphics*[width=14cm]{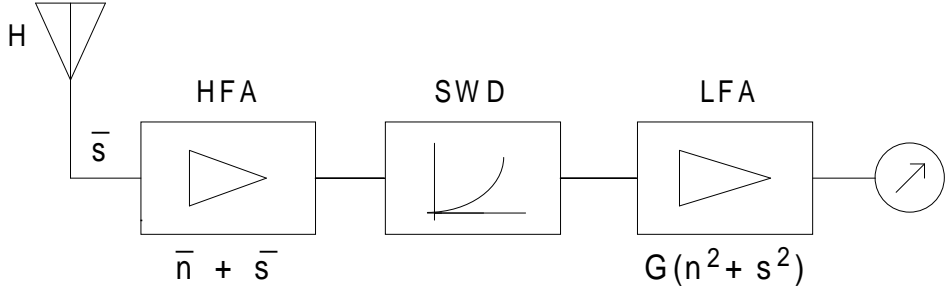}         
 \caption{Block diagram of a radiometer - H = radiation collector, HFA = predetection 
amplifier, SWD = Power detector, LFA = postdetection amplifier,
$\tau$ = integration time, s = signal, n = noise, 
$s^2$ = signal power, $n^2$ = noise power, G = power gain}
          \end{center}
          \label{fig:figure1}            % for cross-references
 \end{figure}

Two classes of noise $n$ are considered here:

a)~{\it white noise} (also called random, or gaussian, or steady state). The 
power spectrum is frequency independent:

\begin{equation}
	w_{wn}(\nu)~d\nu = A_{wn}~d\nu
\label{eq:bianco}
\end{equation}

therefore  

\begin {equation}
      \sigma _{wn}^2~+~\mu_x^2= \left[ A_{wn} (\nu_{max} - \nu_{min})\right]~=~
 A_{wn}~(\frac{1}{\tau} - \frac{1}{T})  
\rightarrow_{N \rightarrow \infty} ~\frac{A_{wn}}{\tau}
\label{eq:sigmabianco}
\end{equation}

No matter which value $\mu_x$ assumes (in many situations $\mu_x = 0$), in
case of reasonable statistics the noise variance approaches quickly a 
constant value. 
In this case the rms fluctuations of the mean value 
decrease as $N$ and $T$ increase.Therefore 
when white noise is dominant one can improve the quality of the data collected 
by a radiometer extending the observing time or increasing the number of independent 
data samples collected. 

b)~$1/f ~noise$. The power spectrum is a power law  
\begin{equation}
	w_{1/f}(\nu)d\nu = {\frac{A_{1/f}}{\nu^{\alpha}}}~d\nu
\label{eq:unosuf}
	\end{equation}

with spectral index $\alpha\sim 1$.    

\[
	\sigma_{1/f}^2 ~+~\mu^2_x= \frac{A_{1/f}}{1~-~\alpha}~[\nu_{max}^{(1-\alpha)} ~-~
\nu_{min}^{(1-\alpha)}] = \frac{A_{1/f}}{1~-~\alpha}~[\tau^{(\alpha -1)} 
~-~ T^{(\alpha -1)}]
\]
\begin{equation} 
~~~~~~~~~\rightarrow _{\alpha \rightarrow 1}~ A_{1/f}~\ln \frac{\nu_{max}}{\nu_{min}} ~ 
=~ A_{1/f}~\ln \frac {T}{\tau}~=~ A_{1/f}~\ln N	
\label{eq:sigmaunosuf}
\end{equation}
 It follows that, when $1/f$ noise is important $(\alpha \ge 1)$, increasing
the observing time or the number of samples collected 
does not help. In fact as T increases  
a growing fraction of low frequency noise is 
added to the system output whose level starts to fluctuates at very low
frequencies. This  effect cannot be cured 
improving the stability of the system temperature or the performance of
the power supply.
\end{section}

\begin{section}{Application to polarimetry}
\begin{subsection}{General layout of a correlation polarimeter}
A common configuration used in radioastronomy for polarimetry is the
two channel correlation receiver shown in figure 2.
\begin{figure}
         \begin{center}
\includegraphics*[width=14cm]{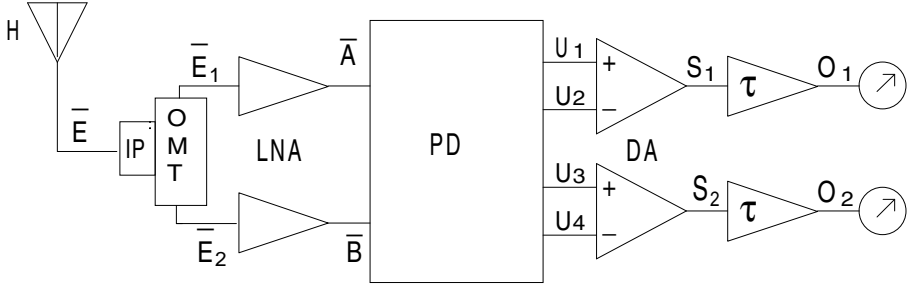}
%\mbox{\epsfig{file=figure2.eps,height=5cm,width=14cm}}   %% Figures
         \end{center}
\caption{Two channel Correlation Receiver - H = Horn, IP = Iris Polarizer, 
OMT = Orthomode transducer, 
LNA = predetection amplifiers, PD = Phase Discriminator, DA = Differential amplifiers,
$\tau$ = post detection amplifiers and integrators, $E, E_1, E_2, A, B$  predetection signals, 
$U_1, U_2, U_3, U_4, S_1, S_2, O_1, O_2$ postdetection signals (see text)}
         \label{fig:figure2}            % for cross-references
\end{figure}
 Fed by a corrugated 
horn, an orthomode transducer (OMT) splits the high frequency signal 
collected by the antenna into orthogonally polarized 
components with a well defined phase difference: $\phi=0$ if the components
are linearly polarized and 
$\phi=\pi/2$ if the components are circularly polarized. 
Inserting or removing an iris polarizer  
between horn and OMT we can set $\phi=\pi/2$ or  
$\phi=0$. The signals available at the OMT outputs are then amplified
by separate receivers and finally injected into the   
Phase Discriminator (PD), a network of four Hybrid circuits and square law detectors
which combines phases and amplitudes of the incoming signals
(\cite{Siro98}, \cite{pever02}). 

To outline the behaviour of the two channel correlation polarimeter we 
go to the frequency domain. If  
$A(\nu)$ and $B(\nu)$ are the monochromatic signals which arrive at the 
inputs of the Phase Discriminator (PD),  
and $\gamma(t) = \phi(t) +\theta$ their phase difference, ($\theta$ is
a constant phase difference which accounts for differences between the electrical 
lengths of the receivers), the PD outputs are:
\begin{equation}
\left [ \matrix{ U_1 \cr U_2 \cr U_3 \cr U_4 \cr} \right ] =
\left [ \matrix{
h_1[ A^2~+~B^2~+~2AB~\cos(\gamma)] \cr h_2[ A^2~+~B^2~-~2AB~\cos(\gamma)] \cr
h_3[ A^2~+~B^2~+~2AB~\sin(\gamma)] \cr h_4[ A^2~+~B^2~-~2AB~\sin(\gamma)] \cr}\right ]
\label{eq:matriceU}
\end{equation}
where $h_i$ describes the overall gain of receivers and PD components.
Differential amplification then gives:
\begin{equation}
\left [ \matrix{S_1 \cr S_2 \cr} \right ] = 
\left [ \matrix{ U_1 - U_2 \cr U_3 - U_4 \cr} \right ] =
\left [ \matrix{
\delta_1 [A^2 + B^2] + \eta_1 [~AB\cos (\gamma)~] \cr
\delta_2 [A^2 + B^2] + \eta_2 [~AB\sin (\gamma)~] \cr} \right ]
\label{eq:matriceS}
\end{equation}
where
\begin{equation}
\left[ \matrix{\delta_1 \cr \delta_2 \cr} \right ] = 
\left [ \matrix{ (h_1 - h_2) \cr (h_3 - h_4) \cr} \right ]
~~~~~~~~~~
\left [ \matrix{\eta_1 \cr \eta_2 \cr} \right ] =
\left [ \matrix{2 (h_1 + h_2) \cr 2 (h_3 + h_4) \cr} \right ]
\label{eq:matriceepiueta}
\end{equation}
Finally integration of $S_1$ and $S_2$ over a time
length $\tau$ gives the outputs $O_1$ and $O_2$. 

For symmetry reasons an ideal receiver should have  
$h_1=h_2, h_3=h_4$ consequently $\delta_i$ should be zero, 
the constant ($\gamma$ independent) terms should vanish and  
$S_1$ and $S_2$ should be sinusoidal functions of $\gamma$ with zero average value. 
Because it is well known (see for instance \cite{Krau}) that  
$<AB\cos (\gamma)>$ and $<AB\sin (\gamma)>$ are linear combinations of the 
Stokes Parameters, we can write:
\begin{equation}
\left [ \matrix{ O_1 \cr O_2 \cr} \right ] = 
\left [ \matrix{ < S_1 > \cr < S_2 > \cr} \right ] =
\left [ \matrix{
K(Q\cos(\theta) - U\sin(\theta)) \cr K(Q\sin(\theta) + U\cos(\theta)) \cr} \right ]
~~~(\phi=\pi/2)
\label{eq:matrice0}
\end{equation}
or
\begin{equation}
\left [ \matrix{ O_1 \cr O_2 \cr} \right ] = 
\left [ \matrix{ < S_1 > \cr < S_2 > \cr} \right ] =
\left [ \matrix{
K(Q\cos(\theta) - V\sin(\theta)) \cr K(Q\sin(\theta) + V\cos(\theta)) \cr} \right ]
~~~ (\phi=0)
\label{eq:stokes}
\end{equation}

The Milano Polarimeter (\cite{Siro98}) is an example of two 
channel correlation polarimeter similar to the one we described above. 
Observations made with two prototypes
(Mk-1 used in 1994 at Baia Terra Nova (Antarctica) and Mk-2 used in 1998 at Dome C 
(Antarctica)) showed  however that in 
spite of the stability and sensitivity provided by the correlation technique, 
both prototypes suffered gain variations and the system outputs had 
offsets (\cite{Siro97}, \cite{Siro98}, \cite{Zann00}). In fact 
small differences between the receiver components give $h_1\simeq h_2$, 
$h_3\simeq h_4$ instead of $h_1 = h_2$ and $h_3 = h_4$  therefore 
the constant terms do not vanish completely  
and offsets of the system outputs appear. Easily these offsets are large compared to 
the amplitude of the sinusoidal terms to be measured. Even worse $1/f$ noise and  
variations of the offset level caused by gain instabilities, mimic signals
produced by polarized sources. 

To cure these effects receivers can be enclosed in a 
(modulation - synchronous detection) loop,a technique widely used  
by radioastronomers (see for instance the 
classical Dicke Receiver (\cite{Krau})). We can modulate the 
power signal or the wave signal, in amplitude or, when applicable, in phase.
Modulation is used to mark the signals we want to detect and
produces a shift of the average receiver output. 
The synchronous detector (also called demodulator) 
picks out only the components of the signals and noise 
marked by modulation and exclude all the remaining components,
above all the noise components, improving the signal to
noise ratio. The demodulator type and the modulation technique must be matched.
 
For polarimetry, where it is essential to preserve phase and amplitude of the
wave signal, phase modulation of the wave signal $E(t)$ is a natural choice, therefore 
the synchronous detector is a Phase Sensitive Detector (PSD). Figure 3
is the block diagram of a Two Channel Correlation Receiver to which Phase Modulation
and Phase Sensitive Detection have been added. Modulator and detector are driven
by a periodic signal whose frequency $\nu_{mod}$ is usually between tens and thousands 
of Hz. 

In the following we analyze the benefits this technique has on the system 
performance.

\begin{figure}
         \begin{center}
\includegraphics*[width=14cm]{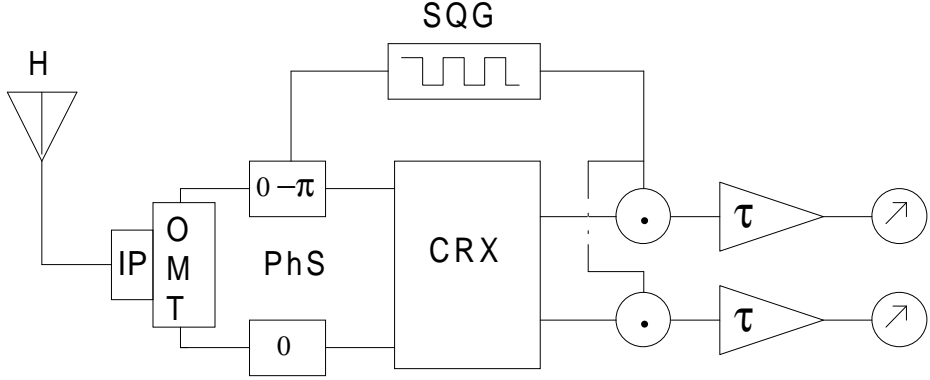}
         \end{center}
\caption{Modulated Correlation Polarimeter -  H = Horn, IP = Iris Polarizer, OMT = 
Orthomode transducer, PhS = Phase shifter, CRX = Correlator, SQG = Square Wave 
Generator, . synchronous detector,
$\tau$ = time integrator}
         \label{fig:figure3b}            % for cross-references
  \end{figure}
 
\end{subsection}
\begin{subsection} {Modulation and Offset Elimination}
Phase modulation of the electric wave $E(t)$ is achieved including into  
arm 1 of the correlation receiver a ($0 - \pi$), just after the OMT, a phase shifter,  
driven by a square wave signal of
period $T_{mod} = 1/\nu_{mod}$ (short compared to $\tau$),
\begin{equation}
r(t) = \left \{
       \begin{array}{cc}
       +1 & ~~~nT_{mod} < t < (n+\frac{1}{2})T_{mod}\\
       -1 & ~~~~~~~(n+\frac{1}{2})T_{mod} < t < (n+1)T_{mod}
       \end{array}
\label{eq:erre}
\right.
\end{equation}
It multiplies $E_1$ by $\pm 1$. An identical phase shifter 
in channel 2, locked in a 
stable position, equalizes the attenuations in
channel 1 and channel 2.
 
If $E_1(t)=E_{o1}(t)e^{i\omega t}$ and $E_2(t)=E_{o2}(t)e^{i(\omega t + \phi)}$, 
are the signals available at the OMT outputs, the inputs of the
Phase Discriminator are
\begin{equation}
\left [ \matrix{ A \cr B\cr} \right ]= 
\left [ \matrix{ k_1 E_1(t) r(t) \cr k_2 E_2(t) \cr }\right ]=
\left [ \matrix{
k_1 E_{o1} e^{i\omega t}~r(t) \cr   
k_2 E_{02} e^{i(\omega t + \phi + \theta)} \cr} \right ]
\label{eq:AB}
\end{equation}
where $k_1$ and $k_2$ account for the system gain between OMT and PD, and

\begin{equation}
\left [ \matrix{ S_1 \cr S_2 \cr} \right ] =
\left [ \matrix{
\delta_1 [|k_1E_1|^2 + |k_2E_2|^2] + \eta_1 k_1k_2
[E_1E_2\cos (\theta + \phi)] r(t) \cr
\delta_2 [|k_1E_1|^2 + |k_2E_2|^2] + \eta_2 k_1k_2
[E_1E_2\sin (\theta + \phi)] r(t)\cr} \right ]
\label{eq:Sr}
\end{equation} 
are the outputs of the differential amplifiers.
 
The Phase Sensitive Detector multiplies $S_1$ and $S_2$ 
by $r'(t)$, a function similar to $r(t)$, and integrates the product 
for a time $\tau$, giving: 
\[
\left [ \matrix{O_1\cr O_2\cr} \right ] =
\left [ \matrix{<S_1\times r'(t)> \cr <S_2\times r'(t)> \cr} \right ]~=~
\]
\[
~~~~~~~~~=~\left [ \matrix{
\delta_1 [~|k_1E_1|^2 + |k_2E_2|^2]<r'(t)> \cr
\delta_2 [~|k_1E_1|^2 + |k_2E_2|^2]<r'(t)> \cr} \right ] ~+~
\]
\begin{equation}
~~~~~~~~~~~~~~~~~~~~~+~\left [ \matrix{ 
\eta_1 k_1k_2 [E_1E_2\cos (\theta + \phi)]<r(t)\times r'(t)> \cr
\eta_2 k_1k_2 [E_1E_2\sin (\theta + \phi)]<r(t)\times r'(t)> \cr } \right ]
\label{eq:Or}
\end{equation}

Because $<r(t)>=0$ and $<r'(t)>=0$ now the offset terms vanish even 
when $\delta_i \ne 0 ~(h_i \ne h_{i+1})$.

Three configurations are possible:

i)~{\it system unlocked}: $r(t)$ and $r'(t)$ are 
generated ~~independently ~~therefore ~$<r(t)\times r'(t)> = 0$

In this condition (marked by apex $ul$)  
\begin{equation}
\left [ \matrix{ O_1^{ul} \cr O_2^{ul} \cr} \right ] =
\left [ \matrix{ 0 \cr 0 \cr} \right ]
\label{eq:Oul}
\end{equation}
always.

ii)~{\it system locked}: PSD and modulator are driven, in phase,  
by the same function $r'(t)\equiv r(t)$ 
(condition marked by apex $l$). 

Now $<r(t) \times r'(t)> 
= <r^2(t)> =$ 1 therefore 
\begin{equation}
\left [ \matrix{O_1^l \cr O_2^l \cr} \right ] =
\left [ \matrix{
\eta_1 k_1k_2<E_1E_2 \cos(\theta + \phi)> \cr 
\eta_2 k_1k_2<E_1E_2 \sin(\theta + \phi)> \cr } \right ]
\label{eq:Ol}
\end{equation}
and one gets the Stokes Parameter (see equations.\ref{eq:matrice0} and \ref{eq:stokes})

iii)~{\it system locked} with a phase difference (time delay $\Delta t$)
between the application of $r(t)$ to the modulator and to the PSD 
($r'(t) = r(t+\Delta t)$, condition marked by apex $\Delta$): 
 
\begin{equation}
\left [ \matrix{O_1^{\Delta} \cr O_2^{\Delta} \cr} \right ] =
\left [ \matrix{
\eta_1 k_1k_2<E_1E_2 \cos(\theta + \phi)r(t) r(t+\Delta t)> \cr 
\eta_2 k_1k_2<E_1E_2 \sin(\theta + \phi)r(t) r(t+\Delta t)> \cr} \right ]
\label{eq:Odelta}
\end{equation}

Because $T_{mod}$ is small compared to $\tau$ and the time during which
the amplitude $E_{01}$ and $E_{02}$ of $E_1$ and $E_2$ are expected to vary, 
we can write 
\[
\left [ \matrix{O_1^{\Delta} \cr O_2^{\Delta} \cr} \right ] \simeq
\left [ \matrix{
\eta_1 k_1k_2<E_1E_2 \cos(\theta + \phi)><r(t) r(t+\Delta t)> \cr
\eta_2 k_1k_2<E_1E_2 \sin(\theta + \phi)><r(t) r(t+\Delta t)> \cr} \right ]~=~ 
\]
\begin{equation} ~~~~~~~~~~= F~
\left [ \matrix{
\eta_1 k_1k_2<E_1E_2 \cos(\theta + \phi)> \cr 
\eta_2 k_1k_2<E_1E_2 \sin(\theta + \phi)> \cr } \right ]
\label{eq:Effe}
\end{equation}
where
\[
F ~=~ <r(t) r(t+\Delta t)>~=~ \frac {~1}{~T} \int_0^T {r(t) r(t+\Delta t) dt}~=~\]
\begin{equation}
 ~~~~~~~~=~~1 ~-~ 4 ~\frac{~|\Delta t|}{T} ~~~~~~(-\frac{~T}{~2}\le\Delta t\le +\frac {~T}{~2}~)
\label{eq:F}
\end{equation}
When $\Delta t \ne 0$ and $\Delta t \ne \pm T/2$, ~$~|F|<1$ and 
a reduction of the system sensitivity occurs.
\end{subsection}

\begin{subsection}{Synchronous Detection and Noise reduction} 

So far we neglected the system noise. Let`s now call ${\varepsilon}_i$  
the noise produced in channel $i$ and $\psi_{ri}(t)$ a random phase.
When we add it to the signal $E_i$ equations \ref{eq:AB} and \ref{eq:Or}
become:
\begin{equation}
\left [ \matrix{ A \cr B \cr} \right ] = 
\left[ \matrix{
k_1 [E_1 r(t) + {\varepsilon}_1e^{i\psi_{r1}(t)}] \cr
k_2 [ E_{2}e^{i(\phi + \theta)} + {\varepsilon}_2 e^{i\psi_{r2}(t)}] \cr} \right ]
\label{eq:ABnoise}
\end{equation}
and 
\[
\left [ \matrix{O_1\cr O_2\cr} \right ]~=~\left [ \matrix{
\delta_1 [~|k_1E_1|^2 + |k_2E_2|^2 + |k_1\varepsilon_1|^2 
                                    + |k_2\varepsilon_2|^2)]<r'(t)> \cr
\delta_2 [~|k_1E_1|^2 + |k_2E_2|^2 + |k_1\varepsilon_1|^2 
                                    + |k_2\varepsilon_2|^2)]<r'(t)> \cr} \right ] ~+~
\]
\[
~~~~~~~~~~~~~+~\left [ \matrix{ 
\eta_1 k_1k_2 [E_1E_2\cos (\theta + \phi)]<r(t)\times r'(t)> \cr
\eta_2 k_1k_2 [E_1E_2\sin (\theta + \phi)]<r(t)\times r'(t)> \cr } \right ]
\] 
\begin{equation}
~~~~~~~~~~~~~~~~~~~~~~~~~~+~\left [ \matrix{
< {\it N}_1 \times r' + {\it N}^{*}_1 > \cr
< {\it N}_2 \times r' + {\it N}^{*}_2 > \cr} \right ]
\label{eq:Olnoise}
\end{equation}
where 

\begin{equation}
\left [ \matrix{ {\it N}_1 \cr {\it N}_2 \cr} \right ] =
2~\left [ \matrix{\delta_1 \cr \delta_2 \cr} \right ]
~k_2^2{\varepsilon}_2 E_2~+~
\left [ \matrix{ \eta_1 \cr \eta_2 \cr} \right ]
~k_1k_2\varepsilon_1(E_2~+~{\varepsilon}_2)
\label{eq:N1}
\end{equation}
and

\begin{equation} 
\left[ \matrix{ {\it N}_1^{*} \cr {\it N}_2^{*} \cr} \right ] = 
2~\left [ \matrix{ \delta_1 \cr \delta_2 \cr}\right ]
~k_1^2\varepsilon_1E_1~+~
\left [ \matrix{ \eta_1 \cr \eta_2 \cr} \right ]
k_1k_2~\varepsilon_2~E_1
\label{eq:N1star}
\end{equation}
are noise terms (${\varepsilon}_1$ and ${\varepsilon}_2$ have random phases). As 
in equation \ref{eq:Or} the offset terms, (which now include $\varepsilon_1^2$ 
and $\varepsilon_2^2$) vanish. 

To appraise the filter action 
of the synchronous detector we have to evaluate the noise variance 
going back to the time domain, (see equations
\ref{eq:sigma}, \ref{eq:fourier} and \ref{eq:sigmammu}). By integration 
of equations \ref{eq:Olnoise}, \ref{eq:N1} and
\ref{eq:N1star} over the frequency bandwith of the receiver 
we get the total power measured at the system output:
\begin{equation}
\left [ \matrix{<V_1(t)> \cr <V_2(t)> \cr} \right ] ~=~ 
\left [ \matrix{<S_1(t) + V_{n,1}(t)\times r(t) + V_{n,1}^* >\cr <S_2(t) + 
V_{n,2}(t)\times r(t) + V_{n,2}^* > \cr} \right ]
\label{eq:Vmedio}
\end{equation}
where $S_i(t)$ is the power of the 
signal, $V_{n,i}(t)$ and $V_{n,i}^*(t)$ are the power of the noise associated 
to $N_i$ and $N_i^*$ respectively. 

We then compute the ratio between the standard deviations of 
the noise calculated when the system is locked and
unlocked:
\begin{equation}
R ~=~ \frac{\sigma_l}{\sigma_{ul}} ~=~
\sqrt{\frac{\sigma^2_{V_{n,i}\times r} ~+~ \sigma^2_{V^*_{n,i}}}
{\sigma^2_{V_{n,i}} ~+~ \sigma^2_{V^*_{n,i}}}}
\label{eq:rapporto}
\end{equation}
Assuming the worst conditions ($1/f$ noise dominant everywhere) 
for the power spectra of $V_{n,i}$ and $V_{n,i}^*$ we set 
$w_{V_{n,i}} = A_i/\nu$ and $w_{V_{n,i}^*} = A_i^*/\nu$. Moreover $k_1\simeq k_2$, 
~${\varepsilon}_1\simeq {\varepsilon}_2\simeq {\varepsilon}$, 
~$E_1\simeq E_2 \simeq E$, ~$N_1\simeq N_2 \simeq N$, ~$N_1^*\simeq N_2^* \simeq N^*$,
~$A_1\simeq A_2 \simeq A$, ~$A_1^* \simeq A_2^* \simeq A^*$, 
~$V_{n,1} \simeq V_{n,2} \simeq V_n$,
~$V_{n,1}^* \simeq V_{n,2}^* \simeq V_n^*$
and $\delta_1 \simeq \delta_2 \simeq \delta 
<< \eta \simeq \eta_1 \simeq \eta_2$. It follows ~$({\it N}^*/{{\it N}}) 
~\simeq ~{E}/(E + \varepsilon) < 1 $ and, when
$E$ is small compared to the system noise, 
(as usual when we look for the fine structures 
of the CMB), $N^* << N$, ~$A >> A^*$ 
and $\sigma^2_{V_n} = A \ln{(T/\tau)} >> 
\sigma^2_{V_n^*} = A^* \ln {(T/\tau)}$.

To get $\sigma_{V_{n,i}\times r}$ first of all we compute 
the power spectrum $w_{V_{n,i}\times r}(\nu)$ of $V_{n,i}\times r$. 
Calculations (see Appendix A and \cite{Spiga}) give (here an in the 
following we omit all $i$ indexes)

	\begin{equation}
	w_{V_n\times r}(\nu) = \frac{4}{\pi^2}\sum_{0}^{\infty}\frac{1}{(2k+1)^2}
\frac{A}{|\nu-\nu_k|} + 
\frac{4}{\pi^2}\sum_{0}^{\infty}\frac{1}{(2k+1)^2}\frac{A}{|\nu+\nu_k|}  
\label{eq:spettroVr}
	\end{equation}

Because $<V_n\times r> = 0$ from ~equation~\ref{eq:sigmammu} follows:
	\[
	\sigma_{V_n\times r}^2 = \int_{\nu_{min}}^{\nu_{max}}w_{V_n\times r}(\nu)\,d\nu 
 = \int_{1/T}^{1/\tau}w_{V_n\times r}(\nu)\,d\nu =
	\]
	\[
 =\frac{4}{\pi^2}\sum_{0}^{\infty}\frac{A}{(2k+1)^2}\int_{1/T}^{1/\tau}
\frac{1}{\nu_k-\nu}\,d\nu + \frac{4}{\pi^2}\sum_{0}^{\infty}\frac{A}{(2k+1)^2}
\int_{1/T}^{1/\tau}\frac{1}{\nu_k + \nu}\,d\nu 
	\]
	\begin{equation}
	= \frac{4A_i}{\pi^2} 
\sum_{0}^{\infty}
\frac{1}{(2k+1)^2}\left[\ln\left(\frac{\nu_k - 1/T}{\nu_k - 1/\tau}\right) +  
\ln\left(\frac{\nu_k + 1/\tau}{\nu_k + 1/T}\right) \right] 
\label{eq:sigmaVr}
	\end{equation}
($1/T \ll 1/\tau \ll \nu_k$ for each $k$). \footnote{When $\nu_{mod} \rightarrow 0$ 
(PSD off) $\sigma_{V\times r}^2 \rightarrow ~A \ln (T/\tau) = \sigma^2_{1/f}$ because	 
$\sum_{0}^{\infty} \frac{1}{(2k+1)^2} = \frac{\pi^2}{8}$}  

Finally neglecting $k>O^3$ terms, we get  

\[		
	R ~=~ \frac{\sigma_{l}}{\sigma_{ul}} =  
\sqrt{\frac{\ln \left(\frac{\displaystyle 
\nu_{mod} - 1/T}{\displaystyle \nu_{mod} -1/\tau}\right)
+ \ln \left(\frac{\displaystyle \nu_{mod} + 1/\tau}{\displaystyle \nu_{mod} +1/T}\right)}
{2\ln\frac{\displaystyle T}{\displaystyle \tau}}}~\simeq
\]
\begin{equation}
~~~~~~~~~~~~~~~~~~~~~~~\simeq~ \sqrt{\frac{(\nu_{max}~-~\nu_{min})/\nu_{mod}}
{\ln (\nu_{max}/\nu_{min})}}
\label{eq:rapportocalcolato}
	\end{equation}
\end{subsection}
\end{section}
\begin{section}*{Discussion}
Above we got formulae which can be used to evaluate 
the noise reduction $R$ or, equivalently, 
the improvement of the system sensitivity $1/R$ one obtains adding 
phase modulation and synchronous detection to a correlation receiver. 
Assuming for instance
$\nu_{mod} = 1 kHz$, $\tau = 5 \:sec ~(\nu_{max}= 0.2 Hz)$ and $T = 60 \:min
~(\nu_{min}=2.8~10^{-4} Hz)$, 
as common in CMB observations, we get $R\simeq 5.5~10^{-3}$ and $1/R \simeq 180$.
In practice the effective reduction can be smaller. In fact:

i)We assumed square wave modulation. It gives maximum efficiency, because
the modulator reaches almost immediately a well defined status, 
and keep it for almost $50\%$ of the modulation cycle. 
However the circuits which carries out this operation must be carefully studied because 
spurious signals are easily triggered by sharp transitions. For that
reason sine wave modulation is sometimes  
preferred. It gives smooth transitions therefore
the generation of spurious signals can be more easily controlled. 
However the signal which drives the modulator varies sinusoidally 
therefore the modulator response can vary and during
an important fraction the modulation cycle be poorly defined.

ii)No matter which shape is preferred, practical modulating functions, 
modulators and detectors are only approximations of the mathematical
functions we assumed. Deviations of the real components  
from their model as well as phase dependence of the 
attenuation and impedance of the modulator and detector may 
produce spurious modulations and/or cycle asymmetries.  
Because when present, they give rise to  
$<r> \neq 0$ and/or $<V \times r> \neq 0$ these effects must be carefully
removed once again through proper design of circuits and devices. Residuals which should 
survive can be cancelled by fine shaping the function which
drives the synchronous detector (e.g. \cite{Siro90})
  
iii)For technical reasons sometimes 
the modulator - demodulator
loop does not include the system front end.
 For instance in the most recent 
model (Mk-3) of the Milano Polarimeter the front end has been set outside the loop 
because no reliable cryogenic phase shifter was available (\cite{Siro01}).
   In this case to analyze the system we have to split gain and noise in channel 1 
in two components: $k_{01}$ and $\varepsilon_{01}$, gain
and noise of the section which precedes the loop, $k_{11}$ and
$\varepsilon_{11}$, gain and noise of the section inside the loop. Now 
~$k_1=k_{01}k_{11}$,~$\varepsilon_1 = \varepsilon_{01} + 
(k_{11}/k_1) \varepsilon_{11}~=~(\alpha + \beta)\varepsilon_1$ 
~~($\alpha=\varepsilon_{01}/\varepsilon_1$,~~
$\beta=(k_{11}/k_1)\varepsilon_{11}/\varepsilon_1$), ~~ 
and equation \ref{eq:ABnoise} becomes
\begin{equation}
\left [ \matrix{ A \cr B \cr} \right ] = 
\left[ \matrix{
k_1 [(E_1 + \varepsilon_{01}e^{i\psi_{r01}})r(t) + k_{11}{\varepsilon}_{11}e^{i\psi_{r1}(t)}] \cr
k_2 [ E_{2}e^{i(\phi + \delta)} + {\varepsilon}_2 e^{i\psi_{r2}(t)}] \cr} \right ]
\label{eq:A1A2Bnoise}
\end{equation}
By an analysis similar to the one we made before, for the noise terms we get now:
\[
\left [ \matrix{ {\it N'}_1 \cr {\it N'}_2 \cr} \right ] =
2~\left [ \matrix{\delta_1 \cr \delta_2 \cr} \right ]
~[ k_1^2{\varepsilon}_{01} E_1~+~k_2^2{\varepsilon}_2 E_2]~+~
\left [ \matrix{\eta_1 \cr \eta_2 \cr} \right ]
~k_{11}k_2\varepsilon_{11}(E_2~+~{\varepsilon}_2)
\]
\begin{equation}
~~~~~~~~~~~~~~~~~~~\simeq~\beta~\left [ \matrix{ {\it N}_1 \cr {\it N}_2 \cr} \right ]~<~ 
\left [ \matrix{ {\it N}_1 \cr {\it N}_2 \cr} \right ] 
\label{eq:Nprimo}
\end{equation}
and

\[
\left[ \matrix{ {\it N'}_1^{*} \cr {\it N'}_2^{*} \cr} \right ] = 
2~\left [ \matrix{ \delta_1 \cr \delta_2 \cr}\right ]
~k_1k_{11}\varepsilon_{11} (E_1~+~{\varepsilon}_{01})~+
\]
\[
~~~~~~~~~~+~\left [ \matrix{ \eta_1 \cr \eta_2 \cr} \right ]
~k_1k_2[E_1\varepsilon_2~+~{\varepsilon}_{01} (E_2~+~\varepsilon_2)]
\]
\begin{equation}
~~~~~~~~~~~~~~~~~~\simeq~\left[ \matrix{ {\it N}_1^{*} \cr {\it N}_2^{*} \cr} \right ]~ 
(1~+~\alpha (1~+~\varepsilon/E)
~>~ \left[ \matrix{ {\it N}_1^{*} \cr {\it N}_2^{*} \cr}) \right ]
\label{eq:Nprimostar}   
\end{equation}
where, to evaluate the approximated expression, we set 
$\varepsilon_1 \simeq \varepsilon_2 \simeq \varepsilon$
and $E_1 \simeq E_2 \simeq E$.  
The noise component ${\it N'}^*$ unaffected
by modulation/demodulation,  is now larger by a factor 
$\simeq 1~+~(\varepsilon_{01}/\varepsilon_1)~+~(\varepsilon_{01}/E_1)$.
~ To keep it small and preserve the
efficiency of the modulation/demodulation process, 
$\varepsilon_{01}$ and $k_{01}$  must be small compared to $\varepsilon_1$ and $k_1$ and,
even more important, possibly free from $1/f$ contribution.
Viceversa the component of the total noise one can control through modulation, $\it{N'}$,
decreases by $\beta$.

iv)Last but not least $\nu_{mod}$ must be far from
harmonics of all periodic signal signals used into the receiver. 
Among them the frequency $\nu_{ac}$ of the AC power supply is particularly dangerous:
it can in fact induce, through residuals ripples on the DC outputs of the 
power supply which feed
amplifiers and active components, modulated signals which are picked up by 
the synchronous detector if $\nu_{mod}$ is close to harmonics of $\nu_{ac}$.
    
All the effects we listed above reduce the efficiency of the modulation / synchronous
detection techniques in improving the stability and in rejecting the noise of
a radiometer. By carefull design of the circuitry which realizes the system
the degradation of R can however be contained sufficiently to make
it a second order effect.

    The above results have been obtained assuming a phase modulated correlation receiver. They 
can be extended to other radiometer configurations, like the classical Dicke Receiver 
(\cite{Krau}) or bolometric systems which use amplitude modulation.
In doing it we must remember that modulation and synchronous detection have different
effects on the performance of a receiver. An important effect    
of modulation is a shift of the average output of the receiver. 
This shift can be used to bring to zero the average
value of the receiver output, making the system insensitive to gain 
fluctuations and allowing large amplifications without saturation. 
Synchronous detection improve the signal to noise ratio, creating  a filter which 
excludes signals not marked by modulation and 
cutting the components of the noise at frequency different from and below $\nu_{mod}$.

Used in the past for classical radioastronomical observations and in
many physical experiments (the so called {\it lock in} technique)
modulation and synchronous detection are today essential to reach the sensitivities 
necessary to study the fine structures of the CMB.
 
\end{section}
\begin{section}*{Acknowledgements}
This work is part of a program aimed at detecting the CMB polarization. It is supported by 
the Italian Ministry for University and Research (COFIN programs), the Italian Program 
for Antarctic Research (CSNA), the National Council for Research (CNR), the University 
of Milano Bicocca and the Italian Space Agency (ASI) 
\end{section}
\appendix
\begin{section}{Appendix A :  Power spectrum of the modulated noise $(V \times r)$}
By proper choice of $t$, the square wave $r(t)$ of period  $T_{mod} = 
1/\nu_{mod}$ can be represented 
by a Fourier serie containing only $\cos$ terms and $\nu_{mod}$ odd multiples 
$\nu_k = (2k+1)\nu_{mod}$:

	\begin{equation}
	r(t)= \frac{4}{\pi}\sum_{k=0}^{+\infty}\frac{(-1)^k}{2k+1}\cos(2\pi\nu_k t)
\label{eq:A1}
	\end{equation}	

 The noise $V$ can be written as a Fourier integral

	\begin{equation}
	V(t)= \int_{-\infty}^{+\infty}a(\nu) e^{i(2\pi\nu t+ \psi_{\nu}(t))} \, d\nu
\label{eq:A2}
	\end{equation}  
where $\psi_{\nu}(t)$ is a randomly  variable phase (here and in the following we will omit 
pedix $i$ which marks the system channel). 
Therefore
      \[	
	V(t) \times r(t)
      =\int_{-\infty}^{+\infty}a(\nu)\cos(2\pi\nu t 
+ \psi_{\nu}(t))\ d\nu \times \ \frac{4}{\pi}\sum_{0}^{\infty}\frac{(-1)^k}{2k+1}
\cos(2\pi\nu_k t)
	\]
\begin{equation}
	+ i \int_{-\infty}^{+\infty}a(\nu)\sin(2\pi\nu t + \psi_{\nu}(t))\ d\nu 
\times \ 
\frac{4}{\pi}\sum_{0}^{\infty}\frac{(-1)^k}{2k+1}\cos(2\pi\nu_k t)
\label{eq:A3}
\end{equation}
Using standard trigonometric formulae we can write
	\[
	V(t) \times r(t) = \frac{2}{\pi}\sum_{k=0}^{+\infty}\frac{(-1)^k}{2k+1} \times 
	\]\[
	\times \bigl\{ \int_{-\infty}^{+\infty} a(\nu)[\cos(2\pi(\nu+\nu_k)t + \psi_{\nu}(t)) + 
\cos(2\pi(\nu-\nu_k)t+\psi_{\nu}(t))]\ d\nu +
	\]
\begin{equation}
	+ i  
\int_{-\infty}^{+\infty} a(\nu)[\sin(2\pi(\nu+\nu_k)t +\psi_{\nu}(t) ) + 
\sin(2\pi(\nu-\nu_k)t)+\psi_{\nu}(t)]\ d\nu \bigl\}
\label{eq:A4}
	\end{equation}
Rearranging terms and omitting for semplicity the time dependence of $\psi$ we get:
	\[
	\int_{-\infty}^{+\infty} a(\nu)[\cos(2\pi(\nu+\nu_k)t +\psi_{\nu}) + 
\cos(2\pi(\nu-\nu_k)t +\psi_{\nu})] d\nu\ =
	\]
	\[
	= \int_{0}^{+\infty} a(\nu)\cos(2\pi(\nu+\nu_k)t +\psi_{\nu}) d\nu + 
\int_{0}^{+\infty} a(\nu) \cos(2\pi(\nu-\nu_k)t+ \psi_{\nu}) d\nu +
	\]
	\[
	+ \int_{-\infty}^{0} a(\nu)\cos(2\pi(\nu+\nu_k)t +\psi_{\nu}) d\nu + 
\int_{-\infty}^{0} a(\nu)\cos(2\pi(\nu-\nu_k)t +\psi_{\nu}) d\nu =
	\]
	\[
	= \int_{+\nu_k}^{+\infty} a(\nu-\nu_k)\cos(2\pi\nu t +\psi_{\nu-\nu_k} ) 
d\nu + \int_{-\infty}^{+\nu_k} a(\nu-\nu_k) \cos(2\pi \nu t +\psi_{\nu-\nu_k} ) d\nu +
	\]
	\[
	+ \int_{-\nu_k}^{+\infty} a(\nu+\nu_k)\cos(2\pi \nu t +\psi_{\nu + \nu_k} ) d\nu 
      \]
      \begin{equation}
~~~~~~~~~~~~+ \int_{-\infty}^{-\nu_k} a(\nu+\nu_k) \cos(2\pi\nu t +\psi_{\nu + \nu_k} ) d\nu
	\label{eq:A5}
	\end{equation}
and 
	\[
	~~~~~\int_{-\infty}^{+\infty} a(\nu)[\sin(2\pi(\nu+\nu_k)t +\psi_{\nu}) + 
\sin(2\pi(\nu-\nu_k)t +\psi_{\nu})] d\nu\ =
	\]
	\[
	= \int_{+\nu_k}^{+\infty} a(\nu-\nu_k)\sin(2\pi\nu t +\psi_{\nu-\nu_k})d\nu  
+ \int_{-\infty}^{+\nu_k} a(\nu-\nu_k)\sin(2\pi\nu t +\psi_{\nu-\nu_k}) d\nu +
	\]
	\[
	+ \int_{-\nu_k}^{+\infty} a(\nu+\nu_k)\sin(2\pi\nu t +\psi_{\nu+\nu_k})d\nu
      \]
      \begin{equation}  
+ \int_{-\infty}^{-\nu_k} a(\nu+\nu_k)\sin(2\pi\nu t +\psi_{\nu+\nu_k}) d\nu 
\label{eq:A6}
	\end{equation}

Therefore
	\[
	V(t) \times \ r(t) =
\frac{2}{\pi}\sum_{k=0}^{+\infty}\frac{(-1)^k}{2k+1} \left[ \int_{+\nu_k}^{+\infty} 
a(\nu-\nu_k) \exp[i(2 \pi\nu t +\psi_{\nu-\nu_k})]d\nu \right. +
	\]
	\[
	+ \int_{-\infty}^{+\nu_k} a(\nu-\nu_k)\exp[i(2 \pi\nu t +\psi_{\nu-\nu_k})] d\nu
      \]\[ 
+ \int_{-\nu_k}^{+\infty} a(\nu+\nu_k) \exp[i(2\pi\nu t +\psi_{\nu+\nu_k})]d\nu + 
	\]
	\begin{equation}
	+ \left.  \int_{-\infty}^{-\nu_k} a(\nu+\nu_k) \exp[i(2\pi\nu t +\psi_{\nu+\nu_k})] d\nu \right] 
\label{eq:A7}	
\end{equation}

Because $\psi_{\nu \pm \nu_k}$ is a random function of $\nu$, the $k$ component 
of the power spectrum of $V\times r$ is a sum of power spectra:
\[	
	w_k(\nu)~=~ \frac{4}{\pi^2(2k+1)^2}\left\{
		\begin{array}{cc}
		|a(-\nu - \nu_k)|^2 & ~~~~~-\infty < \nu < -\nu_k\\
		|a(+\nu + \nu_k)|^2 & ~~~~~-\nu_k < \nu < +\infty
		\end{array}
	\right. ~~+
\]
\begin{equation}
	~~~~~~~~~~~~~~~+
	\left\{
		\begin{array}{cc}
		|a(\nu_k - \nu)|^2 & ~~~~~-\infty < \nu < +\nu_k\\
		|a(\nu - \nu_k)|^2 & ~~~~~+\nu_k < \nu < +\infty
		\end{array}
	\right.
\label{eq:A8}
	\end{equation}
which, when the noise is $1/f$ noise ($|a(\nu)|^2 = |a(-\nu)|^2 = A/|\nu|)$ 
becomes
	
\begin{equation}
	w_k(\nu) = \frac{4}{\pi^2(2k+1)^2}\left[\frac{A}{|\nu-\nu_k|} + 
\frac{A}{|\nu+\nu_k|}\right]  
	\label{eq:A9}
	\end{equation}

Because the phases of the $w_k$ terms vary very rapidly in a random way (they  
are calculated at frequencies different for each $k$), the total power spectrum 
is obtained adding the incoherent terms $w_k$

	\begin{equation}
	w_{V\times r}(\nu) = \frac{4}{\pi^2}\sum_{0}^{\infty}\frac{1}{(2k+1)^2}\frac{A}{|\nu-\nu_k|} + 
\frac{4}{\pi^2}\sum_{0}^{\infty}\frac{1}{(2k+1)^2}\frac{A}{|\nu+\nu_k|}  \label{eq:newspet}
\label{eq:A10}
	\end{equation}
\end{section}

\label{label}

% The phrase \cite{Bai92} produces (Bailyn 1992).
% In the phrase \citeasnoun{Bai95} Bailyn et al. (1995) appear as a noun.
% Affixes (e.g. Barnes et al. 1976) are produced by the phrase
% \citeaffixed{Barnes et al. 1976}{e.g.}.
% Other options of the harvard package, e.g. \citeyear, are not
% reproduced in New Astronomy.

\end{document}